\documentclass[10pt, conference]{IEEEtran}
\usepackage{subfigure}
\usepackage{epsfig,graphicx,psfrag}
\usepackage{amsfonts,amsmath,amssymb}
\usepackage{tikz}

\usetikzlibrary{decorations.pathmorphing} 
\usetikzlibrary{fit}					
\usetikzlibrary{backgrounds}	
\usetikzlibrary{arrows}
\usetikzlibrary{arrows,positioning,fit,backgrounds,shapes}
\usetikzlibrary{calc}

\newtheorem{theorem}{Theorem}[section]

\tikzstyle{arw}=[->,>=latex]
\tikzstyle{node}=[rectangle,draw,outer sep=0pt,minimum width=1.7cm, minimum height=8mm]

\newcommand{\setbr}[1]{{\left\{#1\right\}}}

\begin{document}

\title{Secrecy in Cascade Networks}

\author{
\authorblockN{Paul Cuff -- Princeton University}
}

\maketitle

\begin{abstract}
We consider a cascade network where a sequence of nodes each send a message to their downstream neighbor to enable coordination, the first node having access to an information signal.  An adversary also receives all of the communication as well as additional side-information.  The performance of the system is measured by a payoff function evaluated on actions produced at each of the nodes, including the adversary.  The challenge is to effectively use a secret key to infuse some level of privacy into the encoding, in order thwart the adversary's attempt to reduce the payoff.  We obtain information-theoretic inner and outer bounds on performance, and give examples where they are tight.  From these bounds, we also derive the optimal equivocation for this setting as a special case.
\end{abstract}


\section{Introduction}

This work extends the rate-distortion theory for point-to-point secrecy systems in \cite{schieler-cuff-isit-13} to cascade networks.  Here, an information source (an i.i.d. sequence) is available to the first node in a cascade, and communication is possible among adjacent nodes in the cascade (see Figure~\ref{figure sccs}).  As in \cite{schieler-cuff-isit-13}, after communication, each node produces an action sequence.  Furthermore, an adversary observes all communication as well as additional side-information pertaining to the actions of the nodes and the information source, either causally or non-causally.  The performance of the communication system is measured by average payoff, which is specified by a function of the source, actions of the nodes, and an action produced by the adversary for the purpose of minimizing the payoff.  A limited amount of secret key is available to the communication system (but not the adversary) to enable secrecy.

We give inner and outer bounds on the achievable performance.  Both bounds have properties similar to the main features discovered in the point-to-point setting in \cite{schieler-cuff-isit-13}.  In the point-to-point case the optimal communication system can be understood in two steps, each related to a different notions of coordination defined in \cite{cuff-permuter-cover-10}.  In the first step, empirical coordination is used to send a first layer of information to the receiver.  No attempt is made to keep this information secret from the adversary---accordingly, the secret key is not used in this step.  This is the only information that the adversary gains in addition to the side information.  In the second step, secure strong coordination is performed, conditioned on the first layer of information from step one.

\tikzstyle{block} = [draw, fill=blue!20, rectangle,
    minimum height=2em, minimum width=4em]
\tikzstyle{sum} = [draw, fill=blue!20, circle, node distance=1cm]
\tikzstyle{input} = [coordinate]
\tikzstyle{output} = [coordinate]
\tikzstyle{pinstyle} = [pin edge={to-,thin,black}]

\begin{figure}[ht]
\begin{center}
\resizebox{3in}{!}{
\begin{tikzpicture}[scale=1,auto, node distance=1cm,>=latex']
    \node [input, name=input] {};
    \node [block, below of=input, node distance=1.5cm] (node1) {$F_n$};
    \node [block, right of=node1,
            node distance=3cm] (node2) {$G_n$};
    \node [block, right of=node2,
            node distance=3cm] (node3) {$H_n$};

    \draw [->,double] (node2) -- node[name=r2] {$nR_2$ bits} (node3);
    \draw [] (node3) -- node[name=j2] {$M_2$} (node2);
    \node [output, below of=node2, node distance=1.25cm] (output1) {};
    \node [output, below of=node3, node distance=1.25cm] (output2) {};
    \node [draw=none, above of=node2, node distance = 2cm] (commrand) {$nR_0$ bits};

    \draw [draw,<-] (node1) -- node {$X^n$} (input);
    \draw [->,double] (node1) -- node {$nR_1$ bits} (node2);
    \draw [] (node2) -- node[name=j1] {$M_1$} (node1);
    \draw [->] (node2) -- node [name=yn] {$Y_{(2)}^n$}(output1);
    \draw [->] (node3) -- node [name=zn] {$Y_{(3)}^n$}(output2);
    \draw [->,double] (commrand) -- (node1);
    \draw [->,double] (commrand) -- (node2);
    \draw [->,double] (commrand) -- (node3);

    \node [draw=none, below of=j1, node distance=1.75cm] (evei) {$\setbr{M_1,M_2,W_{(1)}^{t-1},W_{(2)}^{t-1},W_{(3)}^{t-1}}$};
    \node [block, right of=evei, node distance=3.5cm] (eve) {$E_n$};
    \node [draw=none, right of=eve, node distance=1.5cm] (eveo) {$Z_t$};

    \draw [arw] (evei) to node {} (eve);
    \draw [arw] (eve) to node {} (eveo);
\end{tikzpicture}}

\caption{{\em Cascade Source Coding:}  The i.i.d. sequence $X^n$ is given by nature. Messages $M_1$ and $M_2$ satisfy the rate constraints $R_1$ and $R_2$.  Secret key at rate $R_0$ is shared by all three nodes.  The adversary obtains the messages and observes noisy versions of the signals produced by the other nodes but does not have access to the secret key.}
\label{figure sccs}

\end{center}
\end{figure}
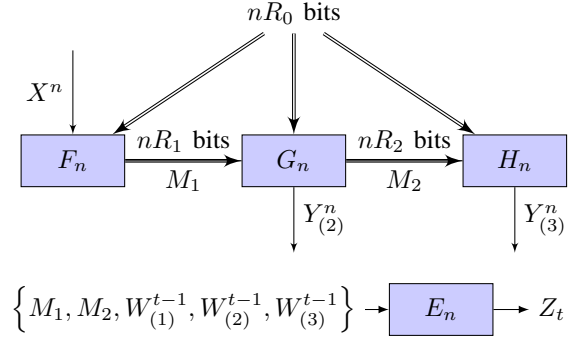

In the cascade setting of this work, we also construct a communication system that designs some information to be recovered by the adversary while keeping the rest perfectly secure.  The secure strong coordination step builds on the result from \cite{satpathy-cuff-isit-13} for the optimal communication design for secure cascade channel synthesis.  We also draw on the optimal empirical coordination design for a cascade network described in \cite{cuff-permuter-cover-10}---the key feature of that result being that the message for the last node in the cascade is the first layer of a superposition code.  The refined layer of the superposition code is sent only to the closer node.

However, the merging of these two separate contexts (i.e. encoding for the cascade network while in the presence of an adversary) presents some subtle conflicts.  We comment on this in Section \ref{subsection interpretation}.

We also examine a couple of special cases of the main result of this work.  One such special case occurs when log-loss functions are used as the payoff function and side-information is chosen appropriately.  This choice allows us to characterize the maximum achievable equivocation measured with respect to the adversary, with distortion constraints at the intended receivers---a familiar metric of secrecy in the literature.

\section{Problem Statement}

\subsection{Communication}

Let $X^n$ be an i.i.d. sequence $X_1,X_2,...,X_n$ referred to as the information source, with distribution $X_1 \sim P_X$.  We consider a sequence of three nodes, with rate-limited communication between them, where the first node observes the source.  The action sequences produced by the other nodes in the network are designated as $Y_{(2)}^n$ for the second node in the cascade and $Y_{(3)}^n$ for the third node.  A uniformly distributed secret key $K \in [2^{nR_0}]$, independent of the information source, is available to all nodes.  The communication system sends messages $M_1 \in [2^{nR_1}]$ from node 1 to node 2 and $M_2 \in [2^{nR_2}]$ from node 2 to node 3 to enable coordination.  The constraints imposed upon the communication system can be summarized in the following way:

\newpage

\noindent
{\em Constraints:}
\begin{enumerate}
\item Source and Key
\begin{eqnarray}
(X^n,K) & \sim & 2^{-nR_0} \prod_{t=1}^n P_X(x_t). \label{constraint independent}
\end{eqnarray}
\item Message Rates
\begin{eqnarray}
M_1 & \in & [2^{nR_1}], \label{constraint M1} \\
M_2 & \in & [2^{nR_2}]. \label{constraint M2}
\end{eqnarray}
\item Markov structure
\begin{eqnarray}
X^n \;\; - & (M_1,K) & - \;\; \left( M_2, Y_{(2)}^n \right), \label{constraint markov 1} \\
\left( X^n, M_1, Y_{(2)}^n \right) \;\; - & (M_2,K) & - \;\; Y_{(3)}^n. \label{constraint markov 2}
\end{eqnarray}
\end{enumerate}

For a given block-length $n$, a rate $(R_0,R_1,R_2)$ communication system is valid if and only if it satisfies the above set of constraints.

\subsection{Adversary's Side information}

As observed in the point-to-point setting of \cite{schieler-cuff-isit-12}, the theory that emerges from this problem is very limited if additional side information at the adversary is not taken into account.  Aside from an overly simplistic theory, two problems result.  The secrecy obtained in the optimally efficient communication system is extremely fragile, meaning that any additional information obtained by the adversary would completely compromise the secrecy design.  Also, with such a limited theory, secrecy results pertaining to equivocation do not occur as a special case.

Thus, in addition to observing the communication $M_1$ and $M_2$, the adversary observes the actions in the network, with or without noise.  We model this side information as outputs of orthogonal memoryless channels from each of the action sequences.  Denoting the collection of all of the side information as $W = \{ W_{(1)}, W_{(2)}, W_{(3)} \}$ we have
\begin{eqnarray}
  \label{eq: side information}
  P_{W|X,Y_{(2)},Y_{(3)}} & = & P_{W_{(1)}|X} P_{W_{(2)}|Y_{(2)}} P_{W_{(3)}|Y_{(3)}}.
\end{eqnarray}

In this work we assume that the side information is observed strictly causally, meaning that the adversary only sees past side information each time he must commit to an action.  This assumption implies that the adversary is forming his reconstruction in real-time and allows for choices of side information channels, such as the identity channel, that would be uninteresting if non-causal observations were available. However, the same communication design is effective even when the side information is non-causal and can be readily characterized.

\subsection{Performance}

A payoff function
\begin{eqnarray*}
  \pi & : & {\cal X} \times {\cal Y}_{(2)} \times {\cal Y}_{(3)} \times {\cal Z} \; \to \; \mathbb{R}
\end{eqnarray*}
encapsulates the objective of the system.  The performance of the system is then measured by
\begin{eqnarray*}
  \bar{\pi} & = & \min_{\{P_{Z_t|{\cal E}_t}\}_{t=1}^n} \mathbb{E} \frac{1}{n} \sum_{t=1}^n \pi(X_t,Y_{(2),t},Y_{(3),t},Z_t).
\end{eqnarray*}
where ${\cal E}_t = \{M_1,M_2,W^{t-1}\}$, which is everything that is available to the adversary at time $t$.

Some other performance metrics stronger than expected average performance yield the same result, as demonstrated in \cite{schieler-cuff-isit-13}.  Furthermore, multiple payoff functions can be optimized jointly, without a significant change to the main result.

\subsection{Achievability}

We say that a rate-payoff tuple $(R_0,R_1,R_2,\Pi)$ is achievable if there exists a block-length $n$ and a valid $(R_0,R_1,R_2)$ communication system that achieves performance $\bar{\pi} \geq \Pi$.

Let ${\cal C}$ be the set of achievable rate-payoff tuples.

\section{Main Result}

\label{section main result}

\subsection{Outer bound}


\begin{theorem}
  \label{thm: outer bound}
  The set of achievable rate-payoff tuples ${\cal C}$ is contained in the set of tuples $(R_0,R_1,R_2,\Pi)$ satisfying
  \begin{eqnarray}
    R_0 & \geq & I(W;V_{(1)}|U), \nonumber \\
    R_1 & \geq & I(X;V_{(1)}), \label{eq:r1 bound} \\
    R_2 & \geq & I(X;V_{(2)}), \label{eq:r2 bound} \\
    \Pi & \leq & \min_{z(\cdot)} \mathbb{E} \; \pi(X,Y_{(2)},Y_{(3)},z(U)), \nonumber
  \end{eqnarray}
  for some distribution $P_{X,Y_{(2)},Y_{(3)},U,V_{(1)},V_{(2)},W}$ satisfying
  \begin{eqnarray}
    X & \sim & P_X, \label{eq:x dist} \\
    P_{W|X,Y_{(2)},Y_{(3)},U,V_{(1)},V_{(2)}} & = & P_{W|X,Y_{(2)},Y_{(3)}} \nonumber \\
    X \;\; - & V_{(1)} & - \;\; Y_{(2)}, \label{eq:v1 chain} \\
    (X,V_{(1)},Y_{(2)}) \;\; - & V_{(2)} & - \;\; Y_{(3)}, \label{eq:v2 chain} \\
    H(V_{(2)},U|V_{(1)}) & = & 0. \nonumber
  \end{eqnarray}
  where $P_{W|X,Y_{(2)},Y_{(3)}}$ is the side information channel stated in \eqref{eq: side information}.  Furthermore, cardinality bounds can be obtained for the variables $U$, $V_{(1)}$, and $V_{(2)}$.
\end{theorem}

\subsection{Inner bound}


\begin{theorem}
  \label{thm:  inner bound}
  The set of achievable rate-payoff tuples ${\cal C}$ contains the set of tuples $(R_0,R_1,R_2,\Pi)$ satisfying \eqref{eq:r1 bound}, \eqref{eq:r2 bound}, and
  \begin{eqnarray*}
    R_0 & > & I(W;V_{(1)}|U_{(1)}), \\
    \Pi & < & \min_{z(\cdot)} \mathbb{E} \; \pi(X,Y_{(2)},Y_{(3)},z(U_{(1)})),
  \end{eqnarray*}
  for some distribution $P_{X,Y_{(2)},Y_{(3)},U_{(1)},U_{(2)},V_{(1)},V_{(2)},W}$ satisfying \eqref{eq:x dist}, \eqref{eq:v1 chain}, \eqref{eq:v2 chain}, and
  \begin{eqnarray*}
    P_{W|X,Y_{(2)},Y_{(3)},U_{(1)},U_{(2)},V_{(1)},V_{(2)}} & = & P_{W|X,Y_{(2)},Y_{(3)}} \\
    H(V_{(2)},U_{(1)}|V_{(1)}) & = & 0, \\
    H(U_{(2)}|U_{(1)}) & = & 0, \\
    H(U_{(2)}|V_{(2)}) & = & 0, \\
    U_{(1)} \;\; - & U_{(2)} & - \;\; V_{(2)},
  \end{eqnarray*}
  where $P_{W|X,Y_{(2)},Y_{(3)}}$ is the side information channel stated in \eqref{eq: side information}.
\end{theorem}

\subsection{Interpretation and Gap}
\label{subsection interpretation}

The variable $U_{(1)}$ in the inner bound (likewise, $U$ in the outer bound) plays the role of information that is designed to be revealed to the adversary in order to reduce the required secret key rate.  Although the inner bound introduces an additional $U_{(2)}$ auxiliary random variables, it is the final Markov relationship in the inner bound, relating the variables $(U_{(1)},U_{(2)},V_{(2)})$, that is the key difference between the bounds.  Its absence would render the two bounds equal.

This Markov chain represents a clash of two conflicting ideas.  In cascade networks, efficient coordination demands first constructing the message for the last node in the network, as the first layer of a superposition code.  On the other hand, in secrecy settings, optimal secrecy is achieved by constructing the public message first.  It is not clear which should get priority in this setting, the public message for the middle node or the private message for the last node---sacrifice secrecy or efficiency.  The Markov chain allows both messages to be constructed at the same layer, without prioritizing one over the other.

\section{Proofs}

\subsection{Outer bound}

Let $T$ be uniformly distributed on the set $\{1,...,n\}$ and independent of all other variables defined thus far with respect to the communication system.

The outer bound involves the following naming of variables:
\begin{eqnarray*}
  X & = & X_T, \\
  Y_{(2)} & = & Y_{(2),T}, \\
  Y_{(3)} & = & Y_{(3),T}, \\
  U & = & \{M_1,M_2,W^{T-1},T\}, \\
  V_{(1)} & = & \{ M_1,M_2,W^{T-1},T,K\}, \\
  V_{(2)} & = & \{ M_2,W_{(1)}^{T-1},W_{(3)}^{T-1},T,K\}, \\
  W & = & W_T.
\end{eqnarray*}
The Markov chains and other conditions of the joint distribution stated in the outer bound can be verified for these variable assignments.

Now we verify the rate and payoff inequalities:
\begin{eqnarray*}
  n R_0 & \geq & H(K) \\
  & \geq & H(K|M_1,M_2) \\
  & \geq & I(W^n;K|M_1,M_2) \\
  & = & \sum_{t=1}^n I(W_t;K|M_1,M_2,W^{t-1}) \\
  & = & n I(W_T;K|M_1,M_2,W^{T-1},T) \\
  & = & n I(W,V_{(1)}|U).
\end{eqnarray*}
\begin{eqnarray*}
  n R_1 & \geq & H(M_1) \\
  & \geq & H(M_1|K) \\
  & \geq & I(X^n;M_1|K) \\
  & = & I(X^n;M_1,K) \\
  & = & I(X^n;M_1,K,M_2) \\
  & = & \sum_{t=1}^n I(X_t;M_1,K,M_2|X^{t-1}) \\
  & = & \sum_{t=1}^n I(X_t;M_1,K,M_2,X^{t-1}) \\
  & = & \sum_{t=1}^n I(X_t;M_1,K,M_2,X^{t-1},Y_{(2)}^{t-1},Y_{(3)}^{t-1}) \\
  & \geq & \sum_{t=1}^n I(X_t;M_1,K,M_2,W^{t-1}) \\
  & = & n I(X_T;M_1,K,M_2,W^{T-1},T) \\
  & = & n I(X;V_{(1)}).
\end{eqnarray*}
The rate $R_2$ is bounded in a similar manner.  Finally,
\begin{eqnarray*}
  \Pi & \leq & \bar{\pi} \\
  & = & \min_{\{P_{Z_t|{\cal E}_t}\}_{t=1}^n} \mathbb{E} \frac{1}{n} \sum_{t=1}^n \pi(X_t,Y_{(2),t},Y_{(3),t},Z_t) \\
  & = & \min_{P_{Z_T|{\cal E}_T,T}} \mathbb{E} \pi(X_T,Y_{(2),T},Y_{(3),T},Z_T), \\
  & = & \min_{P_{Z_T|U}} \mathbb{E} \pi(X,Y_{(2)},Y_{(3)},Z_T), \\
  & = & \min_{z(u)} \mathbb{E} \; \pi(X,Y_{(2)},Y_{(3)},z(U)).
\end{eqnarray*}

\subsection{Inner bound sketch}

The inner bound is proven in the manner of \cite{schieler-cuff-isit-13} and \cite{satpathy-cuff-isit-13}.  Consider the following random codebook construction.  Begin with a codebook ${\cal C}_{U_{(2)}}$ of i.i.d. $u_{(2)}^n$ sequences drawn according to $P_{U_{(2)}}$ and indexed by $m_a \in [2^{nR_a}]$.  For each sequence in ${\cal C}_{U_{(2)}}$ we construct two superposition codebooks.  Let ${\cal C}_{V_{(2)}}$ be independently sampled sequences drawn according to $P_{V_{(2)}|U_{(2)}}(\cdot|u_{(2)}^n(m_a))$ and indexed by the triple $(m_a,m_b,k)$ where $m_b \in [2^{nR_b}]$ and $k \in [2^{nR_0}]$.  Likewise, we construct ${\cal C}_{U_{(1)}}$ according to $P_{U_{(1)}|U_{(2)}}(\cdot|u_{(2)}^n(m_a))$ with sequences indexed by the pair $(m_a,m_c)$ where $m_c \in [2^{nR_c}]$.  Finally, we construct the superposition codebook ${\cal C}_{V_{(1)}}$ according to $P_{V_{(1)}|U_{(1)},V_{(2)}}(\cdot|u_{(1)}^n(m_a,m_c),v_{(2)}^n(m_a,m_b,k))$ with sequences indexed by the tuple $(m_a,m_b,m_c,m_d,k)$ where $m_d \in [2^{nR_d}]$.

To prove achievability, we use the likelihood encoder \cite{cuff-song-itw-13}.  The analysis uses an approximating distribution.  We let all the indices $(M_a,M_b,M_c,M_d,K)$ be uniformly distributed and independent and consider the resulting distributions that arises from applying $(v_{(1)}^n(M_a,M_b,M_c,M_d,K),v_{(2)}^n(M_a,M_c,K))$ as the input to a memoryless channel specified by the conditional distribution $P_{X,Y_{(2)},Y_{(3)}|V_{(1)},V_{(2)}}$.

Define
\begin{eqnarray*}
  M_1 & = & (M_a,M_b,M_c,M_d), \\
  M_2 & = & (M_a,M_b).
\end{eqnarray*}
Furthermore, choose
\begin{eqnarray*}
  R_a = I(X;U_{(2)}) + \epsilon, \\
  R_b = I(X;V_{(2)}|U_{(2)}) + \epsilon, \\
  R_c = I(X;U_{(1)}|V_{(2)}) + \epsilon, \\
  R_d = I(X;V_{(1)}|U_{(1)},V_{(2)}) + \epsilon.
\end{eqnarray*}
The communication constraints in \eqref{constraint M1}, \eqref{constraint M2}, \eqref{constraint markov 1}, and \eqref{constraint markov 2} are satisfied with $R_1 = I(X;V_{(1)}) + 4 \epsilon$ and $R_2 = I(X;V_{(2)}) + 2 \epsilon$.

The remaining part of the proof relies on a soft covering lemma and its extensions (see section VII-B of \cite{cuff-ittrans13}).  It can be shown through iterative steps that the constraint in \eqref{constraint independent} is nearly satisfied in total variation.  Thus, a joint distribution can be constructed (corresponding to the likelihood encoder) which exactly satisfies all of the communication constraints and behaves much like this joint distribution.  Also, it can be shown through similar methods that without knowledge of the key $K$, the posterior distribution of $W^n$ given the messages $(M_1,M_2)$ is nearly that of a memoryless channel from $U_{(1)}^n(M_a,M_c)$ to $W^n$, which renders $W^{t-1}$ useless to the adversary.  After these observations, the inner bound is easy to verify.

\section{Example}

Here we provide an example of a distribution $P_X$ and a payoff function $\pi$ for which the inner and outer bounds are tight and we can characterize the achievable rate-payoff tuples analytically.

Let $P_X$ be the uniform distribution on the set $\{1,2,3\}$, and define the payoff function to be
\begin{eqnarray*}
  \pi(x,y_{(2)},y_{(3)},z) & = & - \infty \; \mathbf{1}_{\{\{x,y_{(2)},y_{(3)}\} = \{1,2,3\}\}} \\
  & & + \; \mathbf{1}_{\{x \neq z\}}.
\end{eqnarray*}
In words, the payoff function is finite if and only if $x$, $y_{(2)}$, and $y_{(3)}$ are all different ternary values, in which case a payoff of 1 is obtained if $z$ differs from $x$.

This unbounded negative payoff function requires a modification to the theorems of Section~\ref{section main result}.  Instead of expected distortion, the average distortion is achieved with high probability.  We don't detail this extension to the theorems in this paper, but it is studied in \cite{schieler-cuff-isit-13}.

For this example, a finite payoff is achievable if and only if\footnote{We ignore the cases of equality, as edge cases, and leave them ambiguous.}
\begin{eqnarray*}
  R_1 & > & \log_2 3, \\
  R_2 & > & \log_2 3 - 1.
\end{eqnarray*}
Assuming the above conditions are met, the optimal payoff $\Pi$ is
\begin{eqnarray*}
  \Pi(R_0) & = & \left\{
  \begin{array}{ll}
    R_0/2, & R_0 \leq 1, \\
    \frac{R_0 - 1}{6(\log_2 3 - 1)} + 1/2, & 1 < R_0 \leq \log_2 3, \\
    2/3, & \log_2 3 < R_0.
  \end{array}
  \right.
\end{eqnarray*}

\subsection{Justification}

To verify, let us first consider the outer bound of Theorem~\ref{thm: outer bound}.  First notice that
\begin{eqnarray*}
  I(X;V_{(1)}) & \geq & I(X;Y_{(2)},Y_{(3)}) \\
  & = & H(X), \\
  I(X;V_{(2)}) & \geq & I(X;Y_{(3)}) \\
  & = & H(X) - H(X|Y_{(3)}),
\end{eqnarray*}
which give the first two inequalities.

The sparsity of the distribution arising from the requirement that $X$, $Y_{(2)}$, and $Y_{(3)}$ all be different with probability one implies that $H(Y_{(3)}|V_{(2)}) = 0$ and $H(Y_{(2)},Y_{(3)}|V_{(1)}) = 0$.  Therefore,
\begin{eqnarray*}
  R_0 & \geq & I(W;V_{(1)}|U) \\
  & = & I(X,Y_{(2)},Y_{(3)};V_{(1)}|U) \\
  & = & H(X,Y_{(2)},Y_{(3)}|U) \\
  & \geq & H(X|U).
\end{eqnarray*}
The optimal tradeoff between $H(X|U)$ and $\min_{z(u)} \mathbb{E} \; \mathbf{1}_{\{X \neq Z\}}$ has been derived elsewhere to be the payoff function $\Pi(R_0)$ stated above.

Now consider the inner bound of Theorem~\ref{thm:  inner bound}.  Construct a joint distribution beginning with $P_{X,Y_{(2)},Y_{(3)}}$ as the uniform distribution over all ternary triples where $X$, $Y_{(2)}$, and $Y_{(3)}$ are unequal.  Also, define $U' = Y_{(2)} - Y_{(3)}$ mod 3.  We focus now on achieving the two corner points of $\Pi(R_0)$, namely, $(R_0,\Pi) = (1,1/2)$ and $(R_0,\Pi) = (\log_2 3,2/3)$.

For the first corner point, let $U_{(2)}$ satisfy the Markov chain $(X,Y_{(2)}) - Y_{(3)} - U_{(2)}$ and be a ternary random variable uniformly distributed on the two values unequal to $Y_{(3)}$.  Let $V_{(2)} = \{Y_{(3)},U_{(2)}\}$ and $U_{(1)} = \{U',U_{(2)}\}$ and $V_{(1)} = \{Y_{(2)},Y_{(3)},U_{(2)}\}$.

For the second corner point, use the same construction but let $U_{(2)} = \emptyset$.

\section{Equivocation}

Traditionally, information theoretic secrecy in the imperfect secrecy regime focusses on maximizing the normalized equivocation of an information source (i.e. $\frac{1}{n} H(X^n|M)$)
while achieving a communication objective.  This approach goes back at least to Wyner's work on the wiretap channel \cite{wyner-wiretap}.  In \cite{cuff-ita13} we showed that the equivocation approach to measuring secrecy is a special case of rate-distortion secrecy with causal disclosure.  In the case of the cascade network, the bounds provided in Theorems \ref{thm: outer bound} and \ref{thm: inner bound} are tight for characterizing equivocation.

The standard metric of equivocation is with respect to the information source $X^n$.  However, in general we can also consider equivocation with respect to the other actions in the network, $Y_{(2)}^n$ and $Y_{(3)}^n$.  For notational simplicity, let us define $${\cal S} \subset \{X,Y_{(2)},Y_{(3)}\}$$ to be the information that we intend to keep secret from the eavesdropper, for which we will characterize the maximum equivocation:
\begin{eqnarray}
  \text{Equivocation } & = & \frac{1}{n} H({\cal S}^n|M_1,M_2).
\end{eqnarray}

We arrive at a characterization of equivocation by using a log-loss payoff function, as has been recently studied as a distortion function in \cite{courtade-wesel} and \cite{courtade-weissman}.

\subsection{Log-loss}
Consider the log-loss payoff function
\begin{eqnarray*}
  \pi (s,z) & = & \log \frac{1}{z(s)}.
\end{eqnarray*}
where $z \in \Delta_{\cal S}$ is a probability mass function.  In other words, the eavesdropper produces a reconstruction $Z_t$ which is a probability distribution over the alphabet of the secured information.  In order to produce the lowest expected distortion, the adversary's best choice of $Z_t$ is the posterior distribution of ${\cal S}_t$ given everything he knows to that point, resulting in an expected payoff that equals conditional entropy.

Specifically, in the case that $W = {\cal S}$ (i.e. the causal side-information is exactly the information that is to be kept secret), we find that the objective $\bar{\pi}$ simplifies as
\begin{eqnarray*}
  \bar{\pi} & = & \min_{ \{P_{Z_t|{\cal E}_t}\}_{t=1}^n } \mathbb{E} \; \frac{1}{n} \sum_{t=1}^n \pi({\cal S}_i,Z_i) \\
  & = & \min_{ \{P_{Z_t|{\cal E}_t}\}_{t=1}^n } \mathbb{E} \; \frac{1}{n} \sum_{t=1}^n \log \frac{1}{Z_i({\cal S}_i)} \\
  & = & \frac{1}{n} \sum_{t=1}^n H({\cal S}_t|M_1,M_2,{\cal S}^{t-1}) \\
  & = & \frac{1}{n} H({\cal S}^n|M_1,M_2) \\
  & = & \text{ Equivocation},
\end{eqnarray*}
where ${\cal E}_t = \{M_1,M_2,{\cal S}^{t-1}\}$.

\subsection{Optimal Equivocation}

We characterize the optimal equivocation $\bar{\pi}$ by applying Theorems~\ref{thm:  outer bound} and \ref{thm: inner bound} with the log-loss payoff function.  In addition to maximizing the equivocation, we assume that the action sequences $Y_{(2)}^n$ and $Y_{(3)}^n$ are constrained to have low distortion with the information source $X^n$ under average distortion metrics
\begin{eqnarray*}
  \mathbb{E} \; \frac{1}{n} \sum_{t=1}^n d_1(X_t,Y_{(2),t}) & \leq & D_1, \\
  \mathbb{E} \; \frac{1}{n} \sum_{t=1}^n d_2(X_t,Y_{(3),t}) & \leq & D_2.
\end{eqnarray*}
This draws on a simple technical extension to Theorems~\ref{thm: outer bound} and \ref{thm:  inner bound}, not detailed in this paper, needed to handle more than one payoff function simultaneously.

\begin{theorem}
\label{thm:  equivocation}
The maximum achievable equivocation of ${\cal S}$ while satisfying distortion contraints $D_1$ and $D_2$ is
\begin{eqnarray*}
  \sup \bar{\pi} & = & \max_{\cal P} \left( H({\cal S}) - \left[ I({\cal S};V_1) - R_0 \right]_+ \right),
\end{eqnarray*}
where $[\cdot]_+ = \max \{0,\cdot\}$ and
\begin{eqnarray*}
  {\cal P} & = & \left\{
  \begin{array}{rcl}
    P_{X,Y_{(2)},Y_{(3)},V_1,V_2} & : & \\
    X & \sim & P_X, \\
    X \;\; - & V_1 & - \;\; Y_{(2)}, \\
    (X,V_1,Y_{(2)}) \;\; - & V_2 & - \;\; Y_{(3)}, \\
    H(V_2|V_1) & = & 0, \\
    \mathbb{E} \; d_1(X,Y_{(2)}) & \leq & D_1, \\
    \mathbb{E} \; d_2(X,Y_{(3)}) & \leq & D_2, \\
    I(X;V_1) & \leq & R_1, \\
    I(X;V_2) & \leq & R_2.
  \end{array}
  \right\}.
\end{eqnarray*}
\end{theorem}

For the case where ${\cal S} = X$, the optimal choices of variables are $V_1 = \{Y_{(2)},Y_{(3)}\}$ and $V_2 = Y_{(3)}$.  Similarly, for both ${\cal S} = \{X,Y_{(2)}\}$ and ${\cal S} = Y_{(2)}$, the optimal choice of $V_2$ is $Y_{(3)}$.

\subsection{Proof of Theorem~\ref{thm:  equivocation}}

The conditions in the definition of ${\cal P}$ come entirely from Theorems~\ref{thm: outer bound} and \ref{thm:  inner bound}.  Let us focus first on the outer bound.  Notice from Theorem~\ref{thm: outer bound},
\begin{eqnarray*}
  R_0 & \geq & I(W;V_{(1)}|U) \\
  & = & I({\cal S};V_{(1)}|U), \\
  \Pi & \leq & \min_{z(u)} \mathbb{E} \; \log \frac{1}{z({\cal S})} \\
  & = & H({\cal S}|U) \\
  & = & H({\cal S}) - I({\cal S};U) \\
  & = & H({\cal S}) - I({\cal S};V_{(1)},U) + I({\cal S};V_{(1)}|U) \\
  & = & H({\cal S}) - I({\cal S};V_{(1)}) + I({\cal S};V_{(1)}|U) \\
  & \leq & H({\cal S}) - I({\cal S};V_{(1)}) + R_0.
\end{eqnarray*}
By identifying $V_1 = V_{(1)}$ and $V_2 = V_{(2)}$ we can conclude that
\begin{eqnarray*}
  \Pi & \leq & H({\cal S}) - \left[ I({\cal S};V_1) - R_0 \right]_+.
\end{eqnarray*}

To prove the inner bound for Theorem~\ref{thm:  equivocation} we construct valid choices of $U_{(1)}$, $U_{(2)}$, $V_{(1)}$, and $V_{(2)}$ and apply Theorem~\ref{thm:  inner bound}.  This is omitted for lack of space.

\section{Acknowledgments}

This work is supported by the NSF (grant CCF-1116013) and the AFOSR (grant FA9550-12-1-0196).

\end{document}